\documentclass[12pt,a4paper]{article}
\usepackage{amssymb} 
\usepackage{graphicx} 
\begin{document}


 \title{ {\bf The strange formula of Dr. Koide}}  
\author{Alejandro Rivero\thanks
       {Zaragoza University at Teruel, {\tt arivero@unizar.es}}
~ and  Andre Gsponer\thanks
       {Independent Scientific Research Institute, Geneva, {\tt gsponer@vtx.ch}}
       }  

\maketitle

\begin{abstract}
We present a short historical and bibliographical review of the lepton mass formula of Yoshio Koide, as well as some speculations on its extensions to quark and neutrino masses, and its possible relations to more recent theoretical developments.
\end{abstract}

\section*{A subjective slice of history}

At the end of 1981 Yoshio Koide, working on some composite models of
quarks and leptons, had the good or bad fortune of stumbling over a very simple relationship between the masses of the three charged leptons \cite[Eq.~(17)]{k1}. This resulted in a prediction of 1.777~GeV/c$^2$ for the mass of the tau lepton. At that time, that prediction was more than two standard deviations away from the measured value, 1.7842~GeV/c$^2$. So by January of 1983 Koide sent to the Physical Review a decaffeinated presentation \cite{k2}, holding a purely phenomenological point of view, and introducing a correction term $\delta$ in order to fit the --- then supposed --- experimental value. Still, this paper paved the way for a nascent research on {\it democratic family mixing}.  This is because Koide mass formula
\begin{equation} 
(m_e+m_\mu+m_\tau) = \frac 23 (\sqrt{m_e}+\sqrt{m_\mu}+\sqrt{m_\tau})^2,
\end{equation}
can be related to the eigenvectors of the well known {\it democratic matrix} 
$$
\pmatrix{1 1 1 \cr 1 1 1 \cr 1 1 1}.
$$

   As Foot pointed out \cite{foot}, Koide's formula is equivalent to ask for
an angle of exactly $\pi/4$ between the eigenvector $(1,1,1)$ and the 
vector formed with the \emph{square roots} of the lepton masses. And the rotation 
around $(1,1,1)$ is of course determined by taking any basis of the nullspace 
of the democratic mixing matrix. But the various uses of this doubly
degenerated matrix are not the theme of this review\footnote{The matrix was first 
used for glue-mediated mixing of mesons. We encourage the reader towards any of the multiple articles from Harald Fritzsch. E.g., \cite{hf} and references therein.}.

   We can only imagine the excitation when some years later the value of the
mass of the tau lepton is revised... And the correction parameter $\delta$ 
becomes plainly zero: The original prediction was right! Koide revives
his formula and builds new models \cite{k3,k4}, but the impact is 
very small.  In the electronic archive {\tt arXiv.org}, R. Foot
\cite{foot} suggests a geometrical interpretation, and a couple years
later Esposito and Santorelli \cite{es} revisited the formula, remarking its
stability under radiative corrections, at least below the electroweak-breaking scale (and breaking at this scale would be, after all, a clue to the origin of the relationship).  Furthermore, it has not escaped that the down quark-family masses are also approximated well enough by the formula, and some effort was made by Koide and others to see how the rest of the particles could fit in the picture.  Contemporary descents of this effort now encompass neutrinos \cite{newone}.

Meanwhile, the increasing interest in seesaw models gave another
opportunity to use the democratic matrix, as well as to generate a top quark mass enhancement justifying a radiative origin for the lower generations.
For some years, seesaw ideas and Koide's formula went hand in hand \cite{k5,k6,k7}. The last incarnation we are aware of, e.g. from Refs.~\cite{k7,k8}, jumps to a justification based on discrete $S(3)$ symmetry. Three-Higgs techniques, used  by several authors in
the late nineties \cite{kiselev,adler} can be given a role there. 

This was only a little cut through the record (see the SPIRES database for more), not the whole story, and for sure it is not over yet.  After all, as it is remarked in \cite{newone}, the current status of the fit against a theoretical unity quotient is $$ 1^{+0.00002635}_{-0.00002021} ~.$$ 

\section*{Form leptons to quarks and neutrinos}

Due to its success with leptons, it was quite obvious that Koide and others would try to extend the mass formula to quarks.  In order to see how this can be done in a physically meaningful way, let us first remove the square roots from Eq.~(1).  Indeed, there is a large consensus that ``mass'' is at least in part due to self-interaction processes which lead to expressions that are quadratic in some underlying classical or quantum field, and thus in the corresponding coupling constant.  For instance, in classical electrodynamics:  $\Delta m = {e^2}/{r}$, in general relativity: $\Delta m = G {m^2}/{r}$, and in field theory:  $\Delta m = g^2 |\Psi|^2$.  This is also the case in pure field theories such as Lanczos's \cite{gh2005} and Weinberg's \cite{weinb}, where mass is entirely originating from the self and mutual interactions of fields. 

   It is therefore natural to introduce a physical quantity that is more fundamental than the mass, which we propose to designate by the symbol $\gimel$, i.e., the third letter in the Hebrew alphabet,\footnote{Being the positional equivalent to ``gamma'', the letter $\gimel$ is called ``gimel.'' It means  ``camel,'' and is thus a nice word for the carrier of the mass.} and to define it as the \emph{gim} of the particle according to the identity $m \equiv \gimel^2$.  This enables to rewrite Koide's formula as
\begin{equation} 
(\gimel_e^2+\gimel_\mu^2+\gimel_\tau^2) = \frac 23 ({\gimel_e}+{\gimel_\mu}+{\gimel_\tau})^2,
\end{equation}
or, by combining the gims of the three lepton into the vector $\vec\gimel = (\gimel_e,\gimel_\mu,\gimel_\tau) \equiv (\sqrt{m_e},\sqrt{m_\mu},\sqrt{m_\tau})$, as the more compact expression
\begin{equation} 
|\vec{\gimel}|^2 = \frac 23 ({\tt Tr~}\vec{\gimel}\,)^2,
\end{equation}
where the trace operation ${\tt Tr}$ acting on a vector is to be interpreted as the sum of its components.  

   In order to apply Eq.~(3) to quarks, as well as to neglect the question of radiative corrections already alluded to in the previous section, one has to circumvent the problems that their masses are not directly measurable, and that their estimated masses are obtained by methods which are different for each generation of them \cite{pdb2004}.

   One possible approach is to use some reasonable model, such as the Nambu-Barut \cite{alfa2,barut} formula generalized by Gsponer and Hurni \cite{gh1995} which gives a smooth and consistent fit to both the lepton and quark data, except for the top quark mass.  Barut's formula for leptons is 
\begin{equation}
  m(N)= m_e (1 + \frac{3}{2} \alpha^{-1} \sum_{n=0}^{n=N} n^4 ),
\end{equation}
where $m_e$ is the mass of the electron and $\alpha \approx 1/137$ the electromagnetic fine structure constant. The masses of the electron, muon, and tau correspond then to $N=0,1,$ and 2, respectively.  The masses of the quarks are also given by this formula, provided $m_e$ is replaced by $m_u  \approx m_e/7.25$, the mass of the $u$-quark. The masses of the $d$, $s$, $c$, and $b$ quarks are then given to an excellent approximation by $N=1,2,3$ and 4, respectively, but the formula fails completely for the $t$ quark.

   While Barut's formula is very different from Koide's, it contains the $3/2$ factor, and a fourth-power dependence on a quantum number $N$ that is typical of self-interaction in non-linear field theories.
There are several ways of using Barut's formula in relations to Koide's:

First, in the generalized Barut formula the charged leptons and three of the quarks are related by a proportionality relation such that $e$ corresponds to $u$, $\mu$ to $d$, and $\tau$ to $s$.  Thus, if this model would be the correct underlying theory of mass, Koide's formula should equally well apply to the $(e,\mu,\tau)$ lepton-triplet, than to the $(u,d,s)$ quark-triplet, i.e., the original Gell~Mann -- Zweig $SU(3)$ triplet.  Indeed, using Barut's masses for either of these triplets, Koide's formula turns out to work at a precision of 2\% for both.  Similarly, if one applies Koide's formula using Barut's theoretical masses for the $(d,s,b)$ triplet (i.e., the $d$-type quark family) the precision is 2.7\%, which is also quite good.  On the other hand for the $(u,c,t)$ triplet (i.e., the $u$-type quark family), the agreement is not as good, i.e., only about 28\%.  Therefore, while Eq.~(1) works rather well for the original quark triplet and the $d$-type quark family, it does not work so well for the $u$-type quark family --- as was observed in Ref.~\cite{newone}, and certainly by Koide and others earlier on.  

A second way is to consider that while there are three massive leptons forming some kind of a three-dimensional real vector, the six quarks could correspond to the six real components of a complex 3-vector.  This picture has the advantage that this complex quark-gim vector could be put in relation to the real lepton-gim vector introduced by Foot \cite{foot}, and that both vectors could have a sound field theoretical interpretation in terms of a real or a complex gim vector field.  Conversely, if the lepton gim-vector is complexified by assuming non-zero neutrino masses, one would have a six partons generalization of Koide's formula.  In that perspective, if the masses of the five first quarks are taken from the generalized Barut formula \cite{gh1995}, and the mass of the $t$ from the data \cite{pdb2004}, i.e., 174 GeV/c$^2$, the precision of the generalized Koide formula, Eq.~(3) in which the sums of the gims and of their squares are taken over all six quarks, is about 9\%.

Finally, a third and possibly the most interesting way is to repeat the previous calculation by taking all six quark masses from the data \cite{pdb2004}.  More precisely, due to the theoretical uncertainties of extracting the quark masses from the data, these six masses are taken as the averages between the extreme values cited in Ref.~\cite{pdb2004}, as listed in Table~1.  Thus, as ${532}^2/180000 = 1.572$ instead of 1.5, Koide's formula applies to a six-quark complex vector with a precision of about 5\%, which is quite good.\footnote{The ratio 283/180 being quite precisely equal to $\pi/2$ illustrates how easily one can get a good agreement with a beautiful number. (Jean-Pierre Hurni, private communication.)}

\begin{table}
\begin{center}
\hskip 0.0cm \begin{tabular}{|c|r|r|}
\hline
\multicolumn{3}{|c|}{\raisebox{+0.2em}{{\bf  \rule{0mm}{6mm} Quark masses and gims}}} \\ 
\hline
\raisebox{+0.2em}{~} \rule{0mm}{6mm} & 
\raisebox{+0.2em}{$m$~~~~~ } & 
\raisebox{+0.2em}{$\gimel$~~~~~~  } \\ 
 
\rule{0mm}{0mm} ~~~ & 
[GeV/c$^2$]                 &
[GeV$^{1/2}$/c]                 \\
\hline
\rule{0mm}{5mm}
u   &        0.5  &    0.71   \\
d   &        6~~~ &   2.5~~   \\
s   &      105~~~ &   10~~~~~ \\
c   &     1250~~~ &   35~~~~~ \\
b   &     4500~~~ &   67~~~~~ \\
t   &   174000~~~ &  417~~~~~ \\
\hline
\rule{0mm}{5mm} total
    &   180000~~~ &  532~~~~~ \\
\hline
\end{tabular}
\end{center}
\caption[Quark masses and gims]{Average of the quark masses given by the Particle Data Book \cite{pdb2004}, and their corresponding gims, i.e., $\gimel \equiv \sqrt{m}$.}    \end{table}

   We can therefore conclude this section by stressing that Koide's formula applies reasonable well to quark masses, either if these masses are calculated with Barut's formula generalized to quarks, or if the masses are taken from the Particle Data Book.  Agreement is particularly good if we take for the masses of the three lighter quarks those given by Barut's formula, or if we take all six quark masses from the data.  This suggests that while Barut's formula seems to contain an important element of truth concerning the light parton masses (where the factors 3/2 and $\alpha \approx 1/137$ appear to play some fundamental role \cite{alfa2}), Koide's formula seems to embody a similar element of truth concerning the heavy partons (including the mass of the top quark which does not fit the Nambu-Barut formula).  However, as is well known, all phenomenological formulas which do not have a completely unambiguous theoretical foundation should be used with caution.  This is illustrated by the less successful fit provided by the $u$-quark triplet as compared to the $d$-quark triplet, which implies that the direct application of Koide's formula to neutrino masses, as is done in Ref.~\cite{newone}, would provide an important clue if it were to be supported by the data.

\section*{Alternate origins of Koide's relation}

{\bf Asymmetric Weyl spinors}. In the early age of single-electron and
pure electrodynamics,
textbooks (such as Berestetskii-Lifschitz-Ditaevskii) told us not to worry about 
different $m$ and $m'$ masses when
combining two Weyl spinors into one Dirac spinor, because their quotient
can always be hidden inside a redefinition of the spinors\footnote{Thus any 
lack of self-adjointness of $m$ was only illusory, wasn't it?}. With the
advent of generations, a new scheme was required to control the new 
freedom of rotation between equally charged particles: The ``CKM'' and ``MNS''
mass-mixing matrices in the quark and lepton sectors. But, is it still true that we can hide from reality any difference in their eigenvalues? 

 Suppose $M$ is a degenerated diagonal matrix $\lambda {\bf 1}_3$, 
and that $M'\neq M^\dagger$ is still a diagonal nondegenerated matrix. The corresponding Dirac 
equation will have a mass squared matrix $MM'$ composed of eigenvalues $\lambda m_i$, and its
real masses will be proportional to $\sqrt m_i$. If one is able to impose some
symmetry in this scheme, Koide's formula will follow, at the cost of three extra
fermions.



{\bf Non-Commutative Geometry}. Models based on non-commutative geometry usually have the
potential of the Higgs sector to be determined by the lepton mass 
matrix, and sometimes by its square and its trace. For instance, in order to get a non-trivial vacuum, the early electroweak Connes-Lott model imposed the condition
\begin{equation}
 3 (m_e^4+m_\mu^4+m_\tau^4) -(m_e^2+m_\mu^2+m_\tau^2)^2 \neq 0 .
\end{equation}
Could this technique be related to Koide's findings?

\section*{Further Remarks}

{\bf Switching couplings off}. It has been remarked and footnoted\footnote{Mostly in the seventies, but there are many
earlier references, e.g., \cite{alfa2, alfa1} and others cited in \cite{gh2005}.} thousands
of times that the muon and the electron (or equivalently the chiral and
electromagnetic breaking scales) are separated by a factor of order
$1/\alpha$. It is less often mentioned \cite{y}, but similarly intriguing, that the
tau and the electroweak vacuum (or equivalently the $SU(3)$ and electroweak scales) are likewise separated by the same Sommerfeld's constant. 

We can mentally
visualize that if we shut down this constant, the mass of the electron
is pushed towards zero, while the Fermi scale is pushed towards infinity.
In this scenario, Foot's hint of an additional symmetry invites to use
Koide's formula as a constraint: The net result is that the muon
mass should increase by a few MeV/c$^2$, while that of the tau should decrease a little in order
to keep the angle. And depending of your views of chiral perturbation theory, the pion should either go to zero mass or just to lose a few MeV/c$^2$ of its mass due
to the component quarks, thus becoming mass degenerated with the
muon!  

Moreover, the switching off of the weak coupling constant will affect
the link between the bottom and the top quarks, causing the former to
lose some weight. In the limit of a very small $\alpha$, we have a bunch of
almost massless particles, another one of hugely massive bosons (completed by the top quark),
and then some surviving elements making use of the $SU(3)$ gap: The muon and 
the strange quark courting themselves and perhaps the pion; the tau, the charmed and bottom quarks dancing around the nucleon
and its glue (see Fig.~1). The mass ratio between tau and muon becomes exactly $({1 + \sqrt{3} \over 1-\sqrt{3}})^2$ i.e., the mass ratio predicted by
Koide's formula when one of the masses (here $m_e$) approaches zero. The {\it three} in this expression is coming from the number of generations, not from the number of colors, and its \emph{numerical value} ($\approx 14$) is very close to the measured coupling constant of the pseudoscalar pion-nucleon theory of strong interactions (see, e.g., \cite[p.450]{deBen}).

One of the puzzling mysteries of Nature is why the massive leptons,
which are colorless, are wandering just there. A hint of universality could
come if we go further, switching off the $SU(3)$ coupling! Even without
contribution from strings and glue, we can build mesons 
starting at the pion scale, because of the strange quark; and baryons at 
the scale of the nucleon, because of the other two massive quarks. Family
masses conspire to save the gap.

\begin{figure}
 \centering
 \includegraphics[width=11cm]{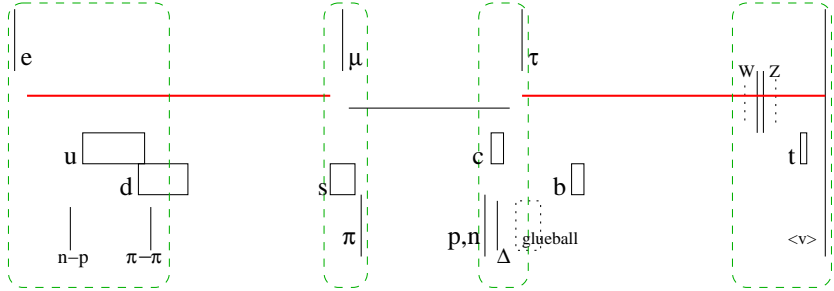}
 \caption{Masses of the elementary particles in logarithmic plot. The
lengths of the two upper horizontal lines, and that of the lower shorter
one, are the numbers $1/\alpha\approx 137$ and $({1 + \sqrt{3} \over
1-\sqrt{3}})^2 \approx 14$, respectively.}
\end{figure}


{\bf Cabibbo angle}. 
From the very beginning \cite{k2}, Koide's formula has been associated to another
one for the Cabibbo angle, involving square roots of the three generations.
Of course when one of the masses is driven to zero, the two extant ones
form a square root of a mass quotient, similar to the kind of expressions 
that nowadays are popular folklore in the phenomenology of mixing. If one 
has followed the {\it gedakenexercise} of the previous paragraph, one will 
 not be surprised that the Cabibbo angle can be obtained both from 
leptons and from quarks. This possibility has also been noticed recently
by Carl Brannen \cite{cb} in a variant, previously used by
Koide \cite{k8}, of the democratic mixing, namely, 
\begin{equation}
 \sqrt{2} \pmatrix{1 &0 &0\cr 0 &1 &0 \cr 0 &0 &1} +
 \pmatrix{ 0 & e^{i\theta} & e^{-i\theta} \cr
           e^{-i\theta} & 0 & e^{i\theta} \cr 
           e^{i\theta} & e^{-i\theta} & 0 },
\end{equation}
which happens to have eigenvalues proportional to our $\sqrt{m_l}$ when
$\theta$ is Cabibbo's angle. In \cite{k8} this solution is avoided,
perhaps intentionally, in order to obtain a complementary fit in the
quark sector.

{\bf Other}. The first author maintains on internet a permanent quest for phenomenologically inspired relationships related to the Standard Model. Please check the wiki page \cite{a1}.


\begin{thebibliography}{30}


\bibitem{k1} Y. Koide, {\it A Fermion-Boson Composite Model of quarks and leptons}, Phys. Lett. B {\bf 120} (1983) 161.
 
\bibitem{k2} Y. Koide, {\it New view of quark and lepton mass hierarchy}, Phys. Rev. D {\bf 28} (1983) 252. 

\bibitem{foot} R. Foot {\it A note on Koide's lepton mass relation}, e-print arXiv:hep-ph/9402242 

\bibitem{hf} H. Fritzsch, {\it Mesons, Quarks and Leptons}, e-print arXiv:hep-ph/0207279

\bibitem{k3} Y. Koide, {\it Charged Lepton Mass Sum Rule from $U(3)$-Family Higgs Potential Model}, Mod. Phys. Lett. A {\bf 5} (1990) 2319--2324.
 
\bibitem{es} S. Esposito and P. Santorelli, {\it A Geometric Picture for Fermion Masses}, Mod. Phys. Lett. A {\bf 10} (1995) 3077-3082, e-print arXiv:hep-ph/9603369 

\bibitem{newone} Nan Li and Bo-Qiang Ma, {\it Estimate of neutrino masses from Koide's relation}, Phys. Lett. B {\bf 609} (2005) 309, e-print arXiv:hep-ph/0505028.

\bibitem{k4} Y. Koide, {\it  New Physics from $U(3)$-Family Nonet Higgs Boson Scenario}, e-print arXiv:hep-ph/9501408.

\bibitem{k5} Y. Koide, {\it Top Quark Mass Enhancement in a Seesaw-Type Quark Mass Matrix}, Z. Phys. C {\bf 71} (1996) 459-468, e-print arXiv:hep-ph/9505201 .

\bibitem{k6} Y. Koide and H. Fusaoka, {\it A Democratic Seesaw Quark Mass Matrix Related to the Charged Lepton Masses}, e-print arXiv:hep-ph/9602303.

\bibitem{k7} Y. Koide, {\it Universal seesaw mass matrix model with an $S_3$ symmetry}, Phys. Rev. D {\bf 60} (1999) 077301, e-print arXiv:hep-ph/9905416 .

\bibitem{k8} Y. Koide, {\it Quark and Lepton Mass Matrices with a Cyclic Permutation Invariant Form}, e-print arXiv:hep-ph/0005137.

\bibitem{kiselev}V.V.Kiselev, {\it Model for three generations of fermions}, 
eprint arXiv:hep-ph/9806523 

\bibitem{adler} S. L. Adler, {\it Model for Particle Masses, Flavor Mixing, and CP Violation Based on Spontaneously Broken Discrete Chiral Symmetry as the Origin of Families}, Phys. Rev. D {\bf 59}(1999) 015012-1--015012-25, e-Print arXiv:hep-ph/9806518 

\bibitem{gh2005} A. Gsponer and J.-P. Hurni, {\it Cornelius Lanczos's derivation of the usual action integral of classical electrodynamics}, Found. Phys. {\bf 35} (2005) 865--880, e-print arXiv:math-ph/0408027.

\bibitem{weinb} S. Weinberg, \emph{A model of leptons}, Phys. Rev. Lett. {\bf 19} (1967) 1264--1266.

\bibitem{pdb2004} S. Eidelman et al., {\it Review of particle physics}, Phys. Lett. B {\bf 592} (2004) 1--1110.

\bibitem{alfa2} Y. Nambu, {\it An empirical mass spectrum of elementary particles}, Prog. Theor. Phys. {\bf 7} (1952) 595--596.

\bibitem{barut} A.O. Barut, {\it Lepton mass formula}, Phys. Rev. Lett. {\bf 42} (1979) 1251.

\bibitem{gh1995} A. Gsponer and J.-P. Hurni, {\it Non-linear  field  theory  for lepton  and quark masses}, Hadronic Journal {\bf 19} (1996) 367--373, e-print arXiv:hep-ph/0201193.

\bibitem{cb} C. Brannen, in internet URL {\tt http://www.physicsforums.com/\\ showpost.php?p=570516\&postcount=112}.

\bibitem{alfa1} K.Matumoto and M.Nakagawa, "Soryushi-ron Kenkyu"
(Particle Physics Research) {\bf 21} (1960). 105

\bibitem{y}  Ray J. Yablon, communicated across internet, in {\tt nntp} bulletin {\tt news:sci.physics.research}

\bibitem{deBen} S. DeBenedetti, Nuclear Interactions (John Wiley \& Sons, New York, 1964) 636.

\bibitem{a1} VV. AA., {\tt http://www.physcomments.org/wiki/}{\tt index.php?}{\tt title=Bakery:HdV}.


\end{thebibliography}
\end{document}